\documentclass[12pt, a4paper, oneside]{article}

\usepackage[latin1]{inputenc}

\usepackage{latexsym,amsmath,amsthm,amsfonts,amssymb}
\usepackage{enumerate,array}
\usepackage{longtable,verbatim,wrapfig}
\usepackage{booktabs,colortbl,setspace}
\usepackage{cancel,anysize}
\usepackage{graphicx}
\usepackage{here}
\usepackage{url}
\usepackage{ragged2e}
\usepackage{multicol}

\usepackage[left=2.5cm,top=2cm,right=2.5cm,bottom=2.5cm]{geometry} 
\usepackage{geometry} 
\usepackage{anysize}
\usepackage{epsfig,graphicx,enumitem}
\usepackage{float}
\usepackage[footnotesize]{caption}
\newcommand{\by}{\boldsymbol{\mathrm{y}}}
\newcommand{\bX}{\boldsymbol{\mathrm{X}}}

\newcommand{\bmp}{\boldsymbol{\mathrm{m}_0}}
\newcommand{\bmpp}{\boldsymbol{\mathrm{m}^*}}
\newcommand{\bth}{\boldsymbol{\theta}}

\newcommand{\bWp}{\boldsymbol{\mathrm{W}_0}}
\newcommand{\bWpp}{\boldsymbol{\mathrm{W}^*}}
\newcommand{\bC}{\boldsymbol{\mathrm{C}}}
\allowdisplaybreaks

\usepackage[absolute]{textpos}

\newtheorem{teo}{Definition}
\usepackage{authblk}
\title{Adaptative significance levels  in linear regression models with known variance}
\author[1]{Alejandra Estefanía Patiño Hoyos}

\author[2]{Victor Fossaluza}

\affil[1]{Institute of Mathematics and Statistics, University of São Paulo, Brazil; alejaeph@ime.usp.br}
\affil[2]{Institute of Mathematics and Statistics, University of São Paulo, Brazil; victorf@ime.usp.br}

\date{}                     
\begin{document}

\maketitle

\begin{abstract}

The Full Bayesian Significance Test (FBST) for precise hypotheses was presented by Pereira and Stern (1999) as a Bayesian alternative instead of the traditional significance test using \textit{p-value}. The FBST is based on the evidence in favor of the null hypothesis (\textbf{H}). An important practical issue for the implementation of the FBST is the determination of how large the evidence must be in order to decide for its rejection. In the Classical significance tests, it is known that \textit{p-value} decreases as sample size increases, so by setting a single significance level, it usually leads \textbf{H} rejection. In the FBST procedure, the evidence in favor of \textbf{H} exhibits the same behavior as the \textit{p-value} when the sample size increases. This suggests that the cut-off point to define the rejection of \textbf{H} in the FBST should be a sample size function. In this work, the scenario of Linear Regression Models with known variance under the Bayesian approach is considered, and a method to find a cut-off value for the evidence in the FBST is presented by minimizing the linear combination of the averaged type I and type II error probabilities for a given sample size and also for a given dimension of the parametric space.
\end{abstract}

\section{Introduction}

The main goal of our work is to determine how small the Bayesian evidence in the FBST should be in order to reject the null hypothesis. Therefore, considering the concepts in  Pereira (1985), in Oliveira (2014) and the recent work of Pereira {\it et al.} (2017) and Gannon {\it et al.} (2019) related to the adaptive significance levels (levels that are function of sample size which are obtained from the generalized form of the Neyman-Pearson Lemma 
), we propose to establish a cut-off value $k^*$ for the $ev\left(\text{\textbf{H}};\by\right)$ as a function of the sample size $n$ and the dimension of the parametric space $d$, i.e., $ k^*= k^*(n,d)$ with $k^*\in [0,1]$, such that $k^*$ minimizes the linear combination of the averaged type I and type II error probabilities, $a\alpha_{\varphi}+b\beta_{\varphi}$.  We will focus on model selection for Linear Regression Models with known variance.

\section{Methodology}

Consider de normal linear regression model 
\begin{equation}\label{model}
\by=\bX\boldsymbol{\theta}+\boldsymbol{\varepsilon}, \quad \boldsymbol{\varepsilon} \sim N_n(\boldsymbol{0},\sigma^2\mathbb{I}_n), 
\end{equation} 
where $\by=(y_1,\dots, y_n)^{\top}$ is an $n \times 1$ vector of $y_i$ observations, $\bX=(\boldsymbol{x}_1,\dots, \boldsymbol{x}_n)^{\top}$
is an $n \times p$ matrix of known coefficients with $\boldsymbol{x}_i=(1,x_{i1},\dots, x_{ip-1})^{\top}$, $\boldsymbol{\theta}=(\bth_{1}^{\top},\bth_{2}^{\top})^{\top}$ is a $p \times 1$ vector of parameters, 
and $\boldsymbol{\varepsilon}=(\varepsilon_1,\dots, \varepsilon_n)^{\top}$ an $n\times1$ vector of random errors. Suppose that the residual error variance $\sigma^2$ is known, then $f(\by\vert \boldsymbol{\theta}) \sim N_n(\bX\boldsymbol{\theta},\sigma^2\mathbb{I}_n)$. The natural conjugate prior family is the family of normal distributions. Suppose therefore that $\bth$ has the $N_p(\bmp,\bWp)$ prior distribution
\begin{equation}\label{likelkv2}
g(\boldsymbol{\theta}) \propto  \exp\left\lbrace -\frac{(\boldsymbol{\theta}-\bmp)^{\top}\bWp^{-1} (\boldsymbol{\theta}-\bmp)}{2} \right\rbrace. 
\end{equation}\\
Then, the posterior distribution of $\bth$ is $\bth \vert \by \sim N_p(\bmpp,\bWpp)$, with
\begin{align}
\bmpp&= (\bWp^{-1}+\sigma^{-2}\bX^{\top}\bX)^{-1}(\bWp^{-1}\bmp+\sigma^{-2}\bX^{\top}\by),\\
\bWpp&=(\bWp^{-1}+\sigma^{-2}\bX^{\top}\bX)^{-1}
\end{align}
If $\bth_1$ has $s$ elements and $\bth_2$ has $r$ elements write
$$\bmp = \left(\begin{array}{c}
\bmp_1\\
\bmp_2\\
\end{array}\right), ~~~~~~
\bWp= \left(\begin{array}
{cc}
\bWp_{11} & \bWp_{12} \\
\bWp_{21}& \bWp_{22} 
\end{array}\right),$$

where $\bmp_1$ is $s\times 1$,  $\bWp_{11}$ is $s\times s$, $\bmp_2$ is $r\times 1$,  $\bWp_{22}$ is $r\times r$. So, 
\begin{eqnarray}
\bth_1 \sim   N_s\left( \bmp_1,\bWp_{11}\right),\qquad
\bth_2 \sim   N_r\left( \bmp_2,\bWp_{22}\right),
\end{eqnarray}

Using general results on multivariate normal distributions, 
\begin{eqnarray}
\bth_1 \vert \bth_2  &\sim &   N_s(\bmp_{1.2}(\bth_2),\bWp_{11.2}),
\end{eqnarray}

where $\bmp_{1.2}(\bth_2)=\bmp_1+\bWp_{12}\bWp_{22}^{-1}(\bth_2-\bmp_2)$ and $\bWp_{11.2}=\bWp_{11}-\bWp_{12}\bWp_{22}^{-1}\bWp_{21}$. A corresponding distribution result if we change $\bmp$ to $\bmpp$ and $\bWp$ to $\bWpp$.
  
\begin{teo}
\begin{sloppypar}
Let $f(\bth\vert \by)$ be the posterior density of $\bth$ given the observed sample. Consider a sharp hypothesis $\text{\textbf{H}}: \boldsymbol{\theta} \in \boldsymbol{\Theta}_\text{\textbf{H}}$ and let ${T_{\by}=\left\lbrace \boldsymbol{\theta} \in \boldsymbol{\Theta}: f(\boldsymbol{\theta} \vert \by)>\text{sup}_\text{\textbf{H}}f(\boldsymbol{\theta} \vert \by) \right\rbrace}$ be the set tangential to \textbf{H}. The measure of evidence in favor \textbf{H} is defined as ${ev\left(\text{\textbf{H}};\by\right)=1-P(\boldsymbol{\theta} \in T_{\by}\vert \by)}$. The \textbf{FBST} is the procedure that rejects \textbf{H} whenever $ev\left(\text{\textbf{H}};\by\right)$ is small (Pereira et al., 2008).
\end{sloppypar}
\end{teo}

Suppose that we want to test the hypotheses
\begin{align}
\text{\textbf{H}}&: \bth_2=\boldsymbol{0}\nonumber\\
\text{\textbf{A}}&: \bth_2\neq \boldsymbol{0}
\end{align}

The tangential set to the null hypothesis is 
\begin{equation}
T_{\by}=\left\lbrace (\bth_1,\bth_2)\in \boldsymbol{\Theta}: f(\bth_1,\bth_2 \vert \by )>\underset{\text{\textbf{H}}}{\operatorname{sup}}f(\bth_1,\bth_2\vert \by ) \right\rbrace, 
\end{equation}
\vspace{10pt}
and, since $(\boldsymbol{\theta}-\bmpp)^{\top}\bWpp^{-1} (\boldsymbol{\theta}-\bmpp) \sim \chi^2_p$, the evidence in favor of \text{\textbf{H}} is 
\begin{equation}
\small
ev\left(\text{\textbf{H}};\by\right)=1-P\left(\chi^2_p<-2\log \left\lbrace \left[\underset{\text{\textbf{H}}}{\operatorname{sup}}f(\bth_1,\bth_2\vert \by)\right] \left|\bWpp\right| ^{1/2}\,(2\pi)^{p/2}\right\rbrace\right),
\end{equation}

where, $\underset{\text{\textbf{H}}}{\operatorname{sup}}f(\bth_1,\bth_2\vert \by)=f(\bmpp_{1.2}(\bth_2=\boldsymbol{0}),\, \boldsymbol{0}\vert \by )$.\\

Consider $\varphi(\by)$ as the test such that

\begin{equation}
\varphi(\by)= \left\{ \begin{array}{l}  0 \quad if \quad ev\left(\text{\textbf{H}};\by\right)> k\\ 1 \quad if \quad ev\left(\text{\textbf{H}};\by\right)\leq 
k.  \end{array} \right.\;
\end{equation}
Thus, define the set
\begin{equation}
\Psi=  \left\lbrace \by \in \boldsymbol{\Omega}: ev\left(\text{\textbf{H}};\by\right)\leq 
k\right\rbrace. 
\end{equation}

The averaged error probabilities can be expressed in terms of the Bayesian prior predictive densities under the respective hypotheses as follows
\begin{align}
\alpha_{\varphi}&= P(\varphi(\by)=1\vert \text{\textbf{H}}) \nonumber\\
&= \int_{\by \in {\Psi}}f_\text{\textbf{H}}(\by)\,d\by\nonumber\\
&=\int_{\by \in {\Psi}}\,\int_{\textbf{H}} f(\by\vert \bth_1,\bth_2) \, g_\textbf{H}(\bth_1,\bth_2)\, d\bth_1\,d\bth_2\nonumber\\
&=\int_{\by \in {\Psi}}\,\int_{\textbf{H}} f(\by\vert \bth_1,\bth_2) \, g(\bth_1\vert \bth_2=\boldsymbol{0})\, d\bth_1\,d\bth_2\\[5pt]
&=\int_{\by \in {\Psi}}\,\int_{\bth_1 \in \mathbb{R}^s} f(\by\vert \bth_1,\bth_2=\boldsymbol{0}) \, g(\bth_1\vert \bth_2=\boldsymbol{0})\, d\bth_1\nonumber\\[5pt]
&=\int_{\by \in {\Psi}}{\small N_n\left( \bX\bC\bmp_{1.2}(\bth_2=\boldsymbol{0}),\,\left(\sigma^2\mathbb{I}_n+(\bX\bC)\bWp_{11.2}(\bX\bC)^{\top}\right)\right),}
\end{align}

where $\bC_{(s+r) \times s}=[\mathbb{I}_s,\boldsymbol{0}_{s\times r}]^{\top}$.
\begin{align}
\beta_{\varphi}&=P(\varphi(\by)=0\vert \text{\textbf{A}})\nonumber\\
&= \int_{\by \notin {\Psi}} f_\text{\textbf{A}}(\by)\,d\by\nonumber\\
&=\int_{\by \notin {\Psi}} \int_{\textbf{A}} f(\by\vert \bth) \, g_\textbf{A}(\bth)\, d\bth\nonumber\\[3pt]
&=\int_{\by \notin {\Psi}} \int_{\textbf{A}} f(\by\vert \bth) \, g(\bth)\, d\bth\nonumber\\[3pt]
&=\int_{\by \notin {\Psi}}  N_n\left( \bX\bmp,\left(\sigma^2\mathbb{I}_n+\bX\bWp\bX^{\top}\right)\right).
\end{align}

So, the adaptive cut-off value $k^{*}$ for $ev\left(\text{\textbf{H}};x\right)$ will be the $k$ that minimizes $a\alpha_{\varphi}+b\beta_{\varphi}$.
\vspace{10pt}

Finally, define $\varphi^{*}(\by)$ as the test such that
\begin{equation}
\varphi^{*}(\by)= \left\{ \begin{array}{l}  0 \quad if \quad ev\left(\text{\textbf{H}};\by\right)> k^{*}\\1 \quad if \quad ev\left(\text{\textbf{H}};\by\right)\leq 
k^{*}.  \end{array} \right.\;
\end{equation}
The optimal averaged error probabilities that depend on the sample size will be
\begin{equation}
\alpha_{\varphi^{*}}^{*}= P(\varphi^{*}(\by)=1\vert \text{\textbf{H}}), \quad \beta_{\varphi^{*}}^{*}=P(\varphi^{*}(\by)=0\vert \text{\textbf{A}}).
\end{equation}

\section{Results}

\begin{figure}[H]
\setlength{\tabcolsep}{-2pt}
\begin{tabular}{cc}
    \includegraphics[scale=0.65]{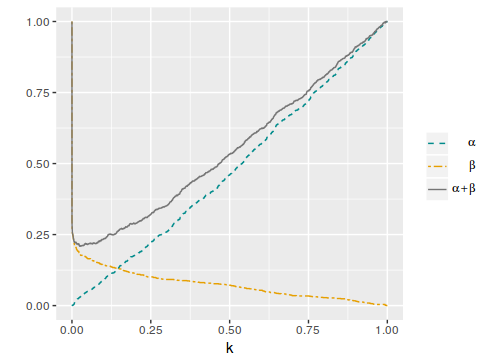} &\includegraphics[scale=0.65]{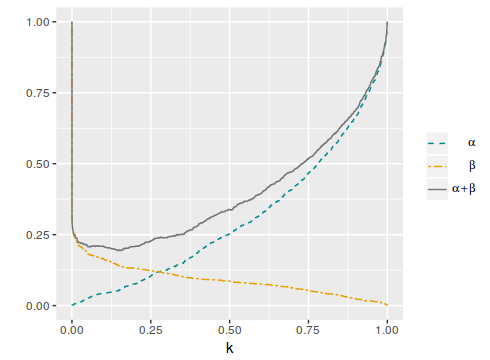} \\
\scriptsize (a) \, $\by=\theta_1+\boldsymbol{\varepsilon}, \,\,  \text{\textbf{H}}:\theta_1=0,$  & \scriptsize(b) \,   $\by=\theta_1+\theta_2 \, x_{i1}+\boldsymbol{\varepsilon}, \,\,  \text{\textbf{H}}:\theta_2=0,$\\
\scriptsize$\mathrm{m}_0=0, \, \mathrm{W}_0=1. $&\scriptsize $\bmp=[0,0]^{\top},\,\, \bWp=\mathbb{I}_2, $ \\[0pt]
 \end{tabular}
\caption{Averaged error probabilities ($\alpha_{\varphi}$, $\beta_{\varphi}$ and $\alpha_{\varphi}+\beta_{\varphi}$) as function of $k$. $n=100$,  $a=b=1$.}
\end{figure}

\setlength\columnsep{5pt}
\begin{multicols}{2}
\begin{table}[H]
\centering
\begin{small}
\begin{tabular}{crc}
  \hline\hline
  &$k^*$  & \\ 
\cline{2-3} 
 $n$  & $d=1$ & $d=2$  \\ 
  \hline 
10 & 0.10040 & 0.35260 \\ 
  50 & 0.05166 & 0.11262 \\ 
  100 & 0.04447 & 0.10473 \\ 
  150 & 0.03776 & 0.09698 \\ 
  200 & 0.03179 & 0.08946 \\ 
  250 & 0.02667 & 0.08226 \\ 
  300 & 0.02244 & 0.07544 \\ 
  350 & 0.01905 & 0.06904 \\ 
  400 & 0.01639 & 0.06311 \\ 
  450 & 0.01429 & 0.05767 \\ 
  500 & 0.01264 & 0.05274 \\ 
  1000 & 0.00649 & 0.02823 \\ 
  1500 & 0.00622 & 0.02954 \\ 
  2000 & 0.00610 & 0.03000 \\ 
   \hline\hline
\end{tabular}
\end{small}
\caption{\small  Cut-off values $k^*$ for $ev\left(\text{\textbf{H}};\by\right)$ as function of $n$, with $d=\text{dim}(\boldsymbol{\Theta})$, $a=b=1$.}
\label{tabknpriorisNN}
\end{table}
\columnbreak
\begin{figure}[H]
\centering
\includegraphics[scale=0.65]{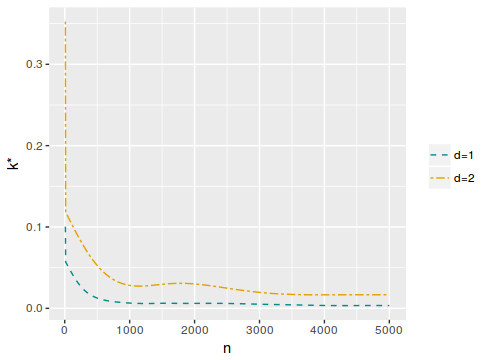}
\caption{ \small Cut-off values $k^*$ for $ev\left(\text{\textbf{H}};\by\right)$ as function of $n$, with $d=\text{dim}(\boldsymbol{\Theta})$, $a=b=1$.}
\label{knpriorisNN}
\end{figure}
\end{multicols}

By increasing $n$, $k^*$ shows a decreasing trend, which means that the influence of sample size on the determination of the cut-off for $ev\left(\text{\textbf{H}};\by\right)$ is very relevant. \\

On the other hand, it is possible to notice the differences in the results between the two models. Then, the cut-off value for $ev\left(\text{\textbf{H}};\by\right)$ will depend not only on the sample size but also on the dimension of the parametric space. More specifically, the $k^*$ value is greater when $d$ is higher.

\begin{figure}[H]
\setlength{\tabcolsep}{-2pt}
\begin{tabular}{cc}
    \includegraphics[scale=0.65]{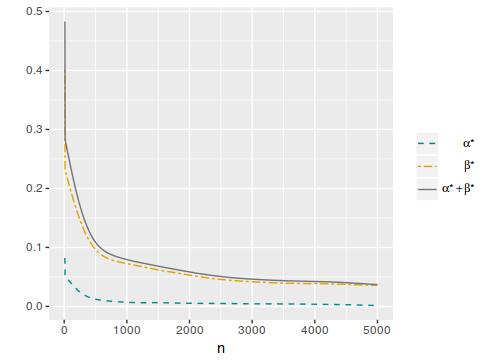} &\includegraphics[scale=0.65]{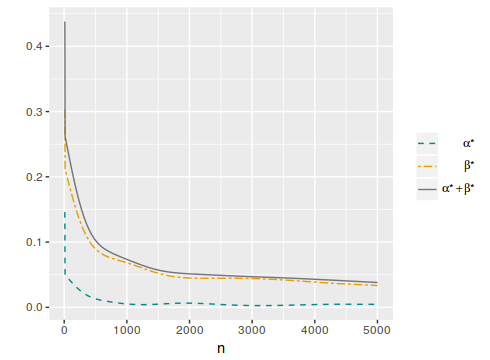} \\
\scriptsize (a) \, $\by=\theta_1+\boldsymbol{\varepsilon}, \,\,  \text{\textbf{H}}:\theta_1=0,$  & \scriptsize(b) \,   $\by=\theta_1+\theta_2 \, x_{i1}+\boldsymbol{\varepsilon}, \,\,  \text{\textbf{H}}:\theta_2=0,$\\
\scriptsize$\mathrm{m}_0=0, \, \mathrm{W}_0=1. $&\scriptsize $\bmp=[0,0]^{\top},\,\, \bWp=\mathbb{I}_2, $ \\[0pt]
\end{tabular}
 \begin{sloppypar} \caption{Optimal averaged error probabilities ($\alpha^{*}_{\varphi^{*}}$, $\beta^{*}_{\varphi^{*}}$ and  $\alpha^{*}_{\varphi^{*}}+\beta^{*}_{\varphi^{*}}$) as function of $n$, $a=b=1$. }  
\end{sloppypar}
\label{kn11}
\end{figure}

With this procedure, increasing the sample size implies that the probabilities of both kind of errors and their linear combination decrease, when in most cases, setting a single level of significance independent of sample size, only type II error probability decreases.

\section*{References}

\begin{description}

{\item[] Gannon, M. A., Pereira, C. A. B. and Polpo, A. (2019).  Blending bayesian and classical tools to define optimal sample-size-dependent significance levels.} {{ \it The American Statistician}, {\bf 73}(sup1), 213-222}

{\item[] Oliveira, M. C. (2014). 
{\it Definição do nível de significância em função do tamanho amostral}. Dissertação de Mestrado, Universidade de São Paulo, Instituto de Matemática e Estatística. Departamento de Estatística, São Paulo.}

{\item[] Pereira, C. A. B., Nakano, E. Y., Fossaluza, V., Esteves, L. G., Gannon, M. A. and Polpo, A. (2017). Hypothesis tests for bernoulli experiments: Ordering the sample space by bayes factors and using adaptive significance levels for decisions. {\it Entropy},  {\bf19}(12), 696.}

{\item[] Pereira, C. A. B., Stern, J. M. and Wechsler, S. (2008). 
an a significance test be genuinely bayesian?. {\it Bayesian Analysis} {\bf 3}(1), 79-100.}

{\item[] Pereira, C. A. B. (1985). 
{\it Teste de hipóteses definidas em espaços de diferentes dimensões: visão Bayesisana e \\interpretação Clássica}. Tese de Livre Docência, Universidade de São Paulo, Instituto de Matemática e Estatística. Departamento de Estatística, São Paulo.}

{\item[] Pereira, C. A. B. and Stern, J. M. (1999). 
Evidence and credibility: Full bayesian significance test for precise hypotheses. {\it Entropy} {\bf 1}(4), 99-110.}

\end{description}

\end{document}